\documentclass{ws-procs975x65}
\usepackage{graphicx}
\usepackage{latexsym}
\usepackage{caption}
\usepackage{amssymb}
\begin{document}
\newcommand{\ti}[1]{\mbox{\tiny{#1}}}
\newcommand{\im}{\mathop{\mathrm{Im}}}
\newcommand{\ap}[1]{\ ^{\mbox{\tiny{(#1)}}}\! }
\title{Effective potential approach to the motion of massive test
particles in Kaluza-Klein gravity}
\author{Valentino Lacquaniti$^1$, Giovanni Montani$^{23}$, Daniela
Pugliese$^{3}$}
\address{$^1$ Physics Department ``E.Amaldi'', University of Rome, ``Roma
Tre"', Via della Vasca Navale 84, I-00146, Rome, Italy
E-Mail: lacquaniti@fis.uniroma3.it\\
$^2$ ENEA-C. R. Frascati Unita' Fus. Mag., via E.Fermi 45,
I-00044, Frascati, Rome, Italy \\
E-Mail: Montani@icra.it\\
$^3$ ICRA—International Center for Relativistic Astrophysics,
Physics Department (G9), University of Rome, ``La Sapienza'',
Piazzale Aldo Moro 5, 00185 Rome, Italy
\\
E-Mail: Pugliese@icra.it}
\begin{abstract}
Effective potential for a class of static solutions of Kaluza-Klein
equations with three-dimensional spherical symmetry is studied. Test
particles motion is analyzed. In attempts to read the obtained
results  with the experimental data, particular attention is devoted
to the Schwarzschild's limit of the four dimensional counterpart of
these electromagnetic  free solutions. Massive particles stable
circular orbits  in particular are studied, and a comparison between
the well known results if the Schwarzschild's case and those found
for the static higher dimensional case is performed. A modification
of the circular stable orbits is investigated in agreement with the
experimental constraints.
\end{abstract}
\section{Generalized Schwarzschild solution}\label{sec:GSS}
Our aim is to analyze  test particles motion  in a Generalized
Schwarzschild solution (GSS) background by an effective potential
approach. First we briefly review the main features of this
solution\cite{Overduin:1998pn,PoncedeLeon:2007bm,Liko:2003ha}. The
metric family
\begin{equation}\label{CGMriunone}
^{5}ds^{2}=\Delta^{\epsilon k}dt^{2}-\Delta^{-\epsilon
(k-1)}dr^{2}-r^{2}\Delta^{1-\epsilon (k-1)}d\Omega^{2}-
\Delta^{-\epsilon }dx^{5\ 2}
\end{equation}
where $\Delta=\left(1-2M/r\right)$ in the 4D-spherical polar
coordinate is a electromagnetic free solution of 5D-Kaluza Klein
equation in the vacuum, with 3D-spherical symmetry. The free
``metric parameters'' $\left(\epsilon,k\right)$ are real constants
related by $ \epsilon^{2}\left(k^{2}-k+1\right)=1 $. Metric
(\ref{CGMriunone}) reduces to the Schwarzschild solution on the
surface $x^{5}=cost$ as $\epsilon\rightarrow0$ and
$k\rightarrow\infty$. In this limit the parameter $M$ is
 the
central body mass. We explored the regions $k\geq0$ and
$\epsilon\geq0$, analyzing  particle motion in  $r>2M$. For these
values of the metric parameters, the GSS solution presents a naked
singularity behavior: it is a black hole one in the Schwarzschild's
limit for $(\epsilon,k)$.
\section{Timelike circular orbits in the GSS
spacetimes.}\label{sec:CircularorGSS} Particles dynamic has been
studied first by  standard approach, considering test particles
motion by a geodesic in 5D-spacetime, then  the analysis has been
performed by an approach \emph{a l$\acute{a}$} Papapetrou, therefore
considering a 5D-particle described by an energy momentum tensor
picked along the particle 4D-world tube. Finally a comparison of the
results obtained into the two different approaches has been made. In
both cases we find the effective potential for the circular polar
orbits of charges as well as neutral test particles, (for the study
of circular orbits in the Schwarzschild case by an effective
potential approach see for example Ref.~\refcite{RuRR}).
\subsection{Geodesic approach}In the first case, assuming a geodesic motion\cite{Kalligas:1994vf,Xu:2007dc}
in 5D-manifold with a constant particle mass $\ap{5}\mu$ the
effective potential reads $\ap{5}V_{eff}\equiv\sqrt{\Delta^{\epsilon
k} \left[1+r^{2}\Delta^{-1+\epsilon (k-1)}\ap{5}L^{2}/\ap{5}\mu^{2}+
\Delta^{\epsilon }\ap{5}\Gamma^{2}/\ap{5}\mu^{2} \right]} $ where
$\ap{5}\Gamma$ is the conserved fifth component of the particle
momentum, $\ap{5}L$ is the conserved quantity associated to the
azimuthal Killing vector $\xi_{\varphi}=\{0,0,0,1,0\}$ per unit rest
mass. The analysis shows that last circular orbits radius,
$r_{lco}=[1+\epsilon(2k-1)]M$ is always located under the expected
values of $r_{lco}=3M$ of the Schwarzschild's limit: circular orbits
(unstable or stable) could be possible also in a region $r<3M$.
\subsection{Papapetrou analysis}
In this section we analyze test particle motion in a GSS background
using the Papapetrou revised approach. This is  realized by an
Papapetrou multipole expansion of a 5D-conservation equation for  an
energy-momentum tensor supposed picked along the 4D-world tube
only\cite{Lacquaniti:2009yy,Lacquaniti:2009rq}. This expansion leads
to different equations of motion for a test particle  in the
4D-spacetime, where the mass $m$ is no more constant but
$\frac{\partial m}{\partial
x^{\mu}}=-\frac{A}{\phi^{3}}\frac{\partial\phi}{\partial x^{\mu}}\,
$ where $A$ is a real function and $\phi^2=-g_{55}$. An effective
potential for a test particle of mass $m$ can be defined as $
\mathfrak{V}_{eff}\equiv\sqrt{g_{00}\left(m^{2}-L^{2}/g_{\varphi\varphi}\right)}
$.  In particular when $A=0$ the equations of motions  describe a
geodetic motion in the ordinary 4D-spacetime for a test particle of
constant mass $m$, where no scalar field coupling term appears.
Formally these are the same equations of motions as those obtained
in the standard geodetic approach for neutral particles. We infer $
r_{c}>\left[1+\epsilon(2k-1)\right]M $ for the circular orbits
radius $r_{c}$. Last stable circular orbit radius is
$r_{lsco}=\left[1+\epsilon(3k-2)+\epsilon\sqrt{(4k-1)(k-1)}\right]M$
where in the Schwarzschild's limit $r_{lsco}=6M$ and $r_{lsco}<6M$
$\forall k>0$. We therefore consider the case $A=cost$, where $m=
A/2\phi^2$ and in the Schwarzschild's limit $m=m_{0}$. Equation of
motion  in this case does not depend on $A$. Last circular orbit  is
located also for this case at $ r_{lco}\equiv
\left[1+\epsilon\left(2k -1\right)\right]M$.  Last stable circular
orbit  is in $$ r_{lsco}/M\equiv\left[\sqrt{4+(15k-8)
\epsilon^{2}-5(3k-8)\epsilon
^{4}}+\epsilon(2+k-11\epsilon+5k\epsilon)+3\right]\left[\epsilon(k+2)\right]^{-1}
$$ this is a free-A quantity, but it is a function of the only
metric parameters $(\epsilon, k)$. Also in this case $r_{lsco}<6M$
and in the Schwarzschild's limit $r_{lsco}=6M$.
 In the case $A=\beta m\phi^{2}$, where $\beta$ is a real number. Mass $m$ follows the
scaling law $m=m_{0}\phi_{0}^{\beta}\phi^{-\beta}$ where
$m_{0}\phi_{0}^{\beta}=cost$ and the equations of motion do not
depend on $m$ but on the constant $\beta$.  For $k>-\beta$ last
circular orbit is located at $r_{lco}\equiv M
\left[1+\epsilon\left(2k +1\right)\right] $. Last stable circular
orbit radius  is in
$$r_{lsco}/M\equiv
\left[(k+2k\beta-3-\beta(2+\beta))\epsilon^2+(k+\beta)\epsilon+3\right]\left[\epsilon(k+\beta)\right]^{-1}
+$$$$+\sqrt{4+\epsilon^{2}\left[-3k(1+2\beta)\left(\epsilon^{2}-1\right)+(2+\beta)\left(\beta-4
+\left(\beta^{3}+2\right)\epsilon
^{2}\right)\right]}\left[\epsilon(k+\beta)\right]^{-1}$$ note that
in the Schwarzschild's limit $r_{lsco}=6M$. Radius of last stable
circular orbit  depends on two parameters, $k$ as the independent
metric parameter and $\beta$ as a ``dynamical'' one. Moreover
$r_{lsco}<6M$ for $\beta>0$, meanwhile for $\beta<0$ and $k>-\beta$,
$r_{lsco}>6M$ is possible. For $\beta=2$ we recover the same
physical situation sketched in the case $A=0$. More generally it is
possible to see that  at an increase of $\beta>0$ for fixed values
of the parameter $k$, an increase of the difference $|r_{lsco}-6M|$
occurs.
\section{Conclusion}
This work was devoted to the study of massive test particle circular
orbits in a GSS solution. We made a comparison between the classical
approach to the dynamic and one based on the Papapetrou multipole
expansion recently applied in a Kaluza Klein context. By an
effective potential approach we found an expression for circular
orbits radius. The stability of such orbits has been investigated
providing the exact last circular orbits radius. As an interesting
result of this analysis we found  that test particles in stable
orbits should exist also for some $r<6M$. This fact, if in agreement
with any experimental observation, could lead to some valuable
constraints to the hypothesi of multidimensional theory in the
Kaluza Klein scenario (see also Ref.~\refcite{Mio1}).

\end{document}